\documentclass{ws-p8-50x6-00}

\def \Rpv{R_{P} \hspace{-1.2em}/\;\hspace{0.2em}}

\begin{document}

\title{Neutrinoless double beta decay and new physics in the neutrino sector}

\author{H.V. Klapdor--Kleingrothaus, H. P\"as}

\address{Max--Planck--Institut f\"ur Kernphysik, P.O. Box 103980, 
D--69029 Heidelberg\\ E-mail: klapdor@gustav.mpi-hd.mpg.de
\\E-mail: Heinrich.Paes@mpi-hd.mpg.de}


\maketitle

\abstracts{
Neutrinoless double beta decay belongs to the most sensitive tools
for the search of new physics beyond the standard model.
The recent half life limit of the Heidelberg--Moscow experiment implies
restrictive bounds on the absolute mass scale in the neutrino sector.
Possible improvements by the GENIUS project provide a unique possibility
to reconstruct the neutrino mass spectrum.
Further constraints on new interactions in the neutrino sector are given in a 
model-independent way. Consequences for neutrino anomalies and theories
beyond the standard model such as left-right symmetric models, R-parity 
violating SUSY and leptoquarks are discussed.
The potential of double beta decay experiments in the search for WIMP dark 
matter is reviewed.}

\section{Introduction}
Double beta decay 
\cite{Kla98,tren} corresponds to two single beta decays 
occuring in 
one
nucleus and 
converts a nucleus (Z,A) into a nucleus (Z+2,A).
While even the standard model (SM) allowed process emitting two antineutrinos
\be
^{A}_{Z}X \rightarrow ^A_{Z+2}X + 2 e^- + 2 {\overline \nu_e}
\ee
is one of the rarest processes in nature with half lives in the region of
$10^{21-24}$ years, more interesting is the search for 
the neutrinoless mode ($0\nu\beta\beta$),
\be        
^{A}_{Z}X \rightarrow ^A_{Z+2}X + 2 e^- 
\ee
which
violates lepton number by two units and thus implies physics beyond the 
SM \cite{ring}. 

The most sensitive experiment so far, the
Heidelberg--Moscow experiment \cite{Kla98,HM99}
is 
searching for the $0\nu\beta\beta$ decay of $^{76}$Ge. 
The results after 31 kg y
measuring time using digital pulse shape analysis correspond to a 
conservative half life limit of \cite{HM99}
\bea
T_{1/2}^{0\nu\beta\beta}&>&1.8 \cdot 10^{25} y ~~~~(90 \% C.L.), \nonumber \\
T_{1/2}^{0\nu\beta\beta}&>&3.0 \cdot 10^{25} y ~~~~(68 \% C.L.).
\eea

To render possible a further breakthrough in search for neutrino masses and 
physics beyond the SM, GENIUS, an experiment operating
a large amount of naked Ge--detectors in a liquid nitrogen shielding,
has been proposed \cite{gen,genbb}.
Operating 288 enriched $^{76}$Ge detectors with a total mass of 1  ton 
inside a nitrogen tank of $\sim$ 12 m height and diameter, one
could access half lifes
of $T_{1/2}^{0\nu\beta\beta}=6 \cdot 10^{27} y$ after one year of measurement. 
A ten ton version would reach a final sensitivity of 
$T_{1/2}^{0\nu\beta\beta}=6 \cdot 10^{29} y$
within 10 years of measurement time.

\begin{figure}
\epsfysize=45mm
\epsfbox{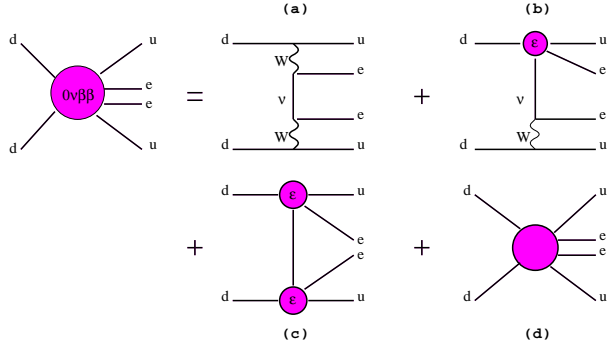}
\caption{\it Feynman graphs of the general double beta rate: 
The contribution a) corresponds to the neutrino mass mechanism with SM 
interactions and is discussed in
the context of neutrino oscillations in  
section 2. The contributions b) - d) include new neutrino interactions and are
discussed in section 3).}
\label{1} 
\end{figure}

\section{Neutrino masses and oscillations}

The search for $0\nu\beta\beta$ decay exchanging a massive 
left--handed Majorana
neutrino between two SM vertices (contribution a) in fig. 1)
at present provides the most 
sensitive approach to determine an absolute neutrino mass and also 
a unique possibility to distinguish between the Dirac or  
Majorana nature of the neutrino.
With the recent half life limit of the Heidelberg--Moscow
experiment \cite{HM99} the following conservative limit on the effective  
neutrino mass 
\bea
\langle m_{\nu}\rangle = \sum_i U_{ei}^2 m_i \leq 0.36 eV 
\hskip5mm (90\% C.L.) \nonumber \\
\langle m_{\nu}\rangle = \sum_i U_{ei}^2 m_i \leq 0.28 eV 
\hskip5mm (68\% C.L.) 
\label{bound}
\eea  
can be deduced. Here the sum extends over light mass eigenstates $m_i$ only.
The GENIUS project could access effective neutrino masses down to $10^{-2}$ eV
or even $10^{-3}$ eV in the 1 ton or 10 ton version, respectively.
Since in specific models of neutrino masses the quantity $\langle m \rangle$
can be related to the oscillation parameters $\Delta m^2$ and 
$\sin^2 2 \theta$,
these bounds imply restrictive bounds on the neutrino mass spectrum, 
which in most cases are more stringent than the bounds 
from precision measurements of the CMB by MAP and Planck.
The extreme cases discussed here are degenerate and hierarchical models
(for a detailed discussion see \cite{smirnov}).

{\it Degenerate models:} Such models have been proposed to get a large mass 
scale for neutrinos acting as hot dark matter and at the same time 
accomodate two of the three neutrino anomalies 
(solar, atmospheric \& LSND neutrinos). In this framework neutrino masses 
up to (few) eV are predicted. Thus large mixing (vacuum or MSW LMA 
oscillations) and strong cancellations
are required to be consistent with eq. \ref{bound}. E.g., maximal 
cancellation, the MSW LMA bestfit and a $\Lambda$CHDM model 
with a Hubble constant 
of $h\simeq 0.5$ implies a value of $ \langle m \rangle =0.15$ eV just below
the present limit.

{\it Hierarchical models:} Such models give less optimistic predictions for 
$0\nu\beta\beta$ decay. However, assuming the MSW LMA solution, still sizable 
contributions up to $\langle m \rangle =$ (few) $10^{-2}$ in the reach of 
GENIUS are possible. Since $0\nu\beta\beta$ decay is most sensitive in the 
large $\Delta m^2$ region of the LMA solution, it may provide 
complementary informations to the search for day-night effects in solar 
neutrinos. 

For a {\it super-heavy left-handed neutrino} a bound of 
$
\langle m_H \rangle = \left(\sum_j \frac{U^2_{ej}}{m_j}\right)^{-1} 
> 9 \cdot 10^7 GeV$
can be deduced,
with heavy mass eigenstates $m_j$. This constraint makes a heavy neutrino 
unobservable at linear colliders except in the most contrived scenarios 
\cite{bel}. 

\section{New interactions}

Besides the exchange of massive Majorana neutrinos between two SM vertices,
a variety of theories beyond the SM predicts new lepton number violating 
interactions contributing to neutrinoless double beta decay,
leading to the idea to construct
the general double beta decay rate allowed by Lorentz--invariance
\cite{Paes98a,supfsh}.
This approach allows to constrain lepton number violating parameters in
arbitrary models. 

For the long range part of the decay rate with two separable vertices
and light neutrino exchange in between (contributions b) and c) in fig. 1), 
one has to consider the Lorentz-invariant contractions of six projections
with defined helicity both for the leptonic ($j_{\alpha}$) and hadronic 
($J_{\alpha}$) current.
The general Lagrangian can be written in terms of
effective couplings $\epsilon^{\alpha}_{\beta}$, which correspond to the
pointlike vertices at the Fermi scale so that Fierz rearrangement is 
applicable:
\be
{\cal L}=\frac{G_F}{\sqrt{2}}\{
j_{V-A}^{\mu}J^{\dagger}_{V-A,\mu}+ \sum_{\alpha,\beta} 
 ^{'}\epsilon_{\alpha}^{\beta}j_{\beta} J^{\dagger}_{\alpha}\}
\ee
with the combinations of hadronic and leptonic Lorentz currents of 
defined helicity 
$\alpha,\beta=V-A,V+A,S-P,S+P,T_L,T_R$.
The prime indicates that
the sum runs over all contractions allowed by 
Lorentz--invariance,
 except for $\alpha=\beta=V-A$.
Here $\epsilon_{\alpha}^{\beta}$ denotes the 
strength of the non--SM couplings. 
For the helicity suppressed terms proportional to the (from below)
unconstrained neutrino mass no limit can be derived and terms proportional
$(\epsilon_{\alpha}^{\beta})^2$ can be neglected. 
The limits on the remaining non--SM couplings derived in s-wave 
approximation and evaluated ``on axis'' are \cite{Paes98a}
(here and in the following $90 \% C.L.$):
$\epsilon^{V+A}_{V+A}< 6 \cdot 10^{-7}$, 
$\epsilon^{V+A}_{V-A}< 4 \cdot 10^{-9}$,
$\epsilon^{S+P}_{S+P}< 9 \cdot 10^{-9}$,
$\epsilon^{S+P}_{S-P}< 9 \cdot 10^{-9}$,
$\epsilon^{T_{R}}_{T_{R}}< 1 \cdot 10^{-9}$,
$\epsilon^{T_{R}}_{T_{L}}< 6 \cdot 10^{-10}$.
These bounds e.g. exclude the possibility to fake the LSND anomaly 
(and this way accomodate for all neutrino anomalies with only three neutrinos)
via the 
lepton number violating reaction $\nu_e u_L \rightarrow d_R e^+$ in a 
model-independent way \cite{bergm}.

For the short range part the hadronic currents have to be contracted with
leptonic currents $j_{\alpha}=\bar{e}{\cal O}_{\alpha}e^C$, where 
${\cal O}_{\alpha}$
denotes the operators of defined helicty discussed above. In this case
the general Lagrangian is
\bea
{\cal L}= \frac{G_F^2}{2} m_P^{-1} \{ 
\epsilon_1 JJj 
+ \epsilon_2 J^{\mu\nu}J_{\mu\nu}j
+ \epsilon_3 J^{\mu}J_{\mu}j 
+ \epsilon_4 J^{\mu}J_{\mu\nu}j^{\nu}
+ \epsilon_5 J^{\mu}J j_{\mu} \nonumber \\
+ \epsilon_6 J^{\mu}J^{\nu}j_{\mu\nu}
+ \epsilon_7 JJ^{\mu\nu}j_{\mu\nu}
+ \epsilon_8 J_{\mu\kappa}J^{\nu\kappa}j^{\mu}_{\nu}\},
\eea
where indices $\alpha$ have been suppressed. Since no fundamental tensors 
exist in renormalizable theories and since the leptonic tensor current vanishes
in the s-wave approximation, the contributions proportional to
$\epsilon_4$, $\epsilon_6$, $\epsilon_7$, $\epsilon_8$ can be neglected. 
The remaining terms are constrained as follows \cite{supfsh}: 
$\epsilon_1<3 \cdot 10^{-7}$, 
$\epsilon_2<2 \cdot 10^{-9}$, 
$\epsilon_3<4 \cdot 10^{-8}/1 \cdot 10^{-8}$ 
($V\mp A~ V \mp A /  V\mp A ~V \pm A$), 
$\epsilon_5<2 \cdot 10^{-7}$.

\section{Left-right symmetry, R-parity violation, Leptoquarks}

In this section we apply the general discussion above to specific theories of
physics beyond the SM.

{\it Left--Right--Symmetric Models:}
In left--right symmetric models the left--handedness of weak 
interactions is explained as due to the effect of different symmetry breaking 
scales in the left-- and in the right--handed sector. 
$0\nu\beta\beta$ decay proceeds through exchange of the heavy right--handed 
partner of the ordinary neutrino between right-handed W vertices, leading
to a limit of
\be
m_{W_{R}}\geq 1.4 \Big(\frac{m_N}{1TeV}\Big)^{-(1/4)} TeV.
\ee 
Including a theoretical limit obtained from considerations of vacuum 
stability \cite{moha86} one can deduce an absolute lower limit on the 
right--handed W mass of \cite{11} 
\be
m_{W_{R}}\geq 1.4 TeV.
\ee

{\it Supersymmetry:}
While in the minimal supersymmetric extension (MSSM) R--parity
is assumed to be conserved, there are no theoretical reasons for $R_p$ 
conservation and
several GUT and Superstring models require
R--parity violation in the low energy regime.  
In this case $0\nu\beta\beta$ decay can occur through Feynman graphs 
involving the exchange of
superpartners as well as $\Rpv$--couplings $\lambda^{'}$ 
\cite{hir95,hir95d,hir96c,hir96,Paes98}.
The half--life limit of the Heidelberg--Moscow experiment leads to bounds
in a multidimensional parameter space \cite{hir95,hir96c}
\be
\lambda_{111}^{'}\leq 4\times 10^{-4}\Big(\frac{m_{\tilde{q}}}{100 
GeV} \Big)^2
\Big(\frac{m_{\tilde{g}}}{100 GeV} \Big)^{1/2}
\ee
(for $m_{\tilde{d}_{R}}=m_{\tilde{u}_{L}}$), which are the sharpest limits on
 $\Rpv$--SUSY.

$0\nu\beta\beta$ decay is not only sensitive to 
$\lambda_{111}^{'}$. Taking into account the fact that the SUSY partners of the
left- and right--handed quark states can mix with each other, new diagrams 
appear in which the neutrino-mediated double beta decay is accompanied by
SUSY exchange in the vertices \cite{babu95,hir96}.
A calculation of previously neglected tensor contributions to the decay rate
allows to derive improved limits on different combinations of $\lambda^{'}$
\cite{Paes98}. 
Assuming the supersymmetric mass parameters of order 100 GeV, the half life 
limit of the Heidelberg--Moscow Experiment implies:
$\lambda_{113}^{'} \lambda_{131}^{'}\leq 3 \cdot 10^{-8}$,
$\lambda_{112}^{'} \lambda_{121}^{'}\leq  1 \cdot 10^{-6}$

In addition, stringent bounds on coupling products can be derived directly
from the effective mass bound eq. \ref{bound}, since R-parity violating 
interactions will produce neutrino Majorana masses on loop level.
It implies \cite{bhat}
$\lambda^{'}_{133} \lambda^{'}_{133} < 5 \cdot 10^{-8}$,
$\lambda^{'}_{132} \lambda^{'}_{123} < 1 \cdot 10^{-6}$,
$\lambda^{'}_{122} \lambda^{'}_{122} < 3 \cdot 10^{-5}$,
$\lambda_{133} \lambda_{133} < 9 \cdot 10^{-7}$,
$\lambda_{132} \lambda_{123} < 2 \cdot 10^{-5}$,
$\lambda_{122} \lambda_{122} < 2 \cdot 10^{-4}$.

In the case of R--parity conserving SUSY, based on a theorem proven in 
\cite{sneut},
the $0\nu\beta\beta$ mass limits can be converted in sneutrino Majorana mass
term limits being more restrictive than what could be obtained
in inverse neutrinoless double beta 
decay and single sneutrino production at future linear colliders (NLC)
\cite{sneut}.
 
{\it Leptoquarks:}
Leptoquarks are scalar or vector particles coupling both to leptons 
and quarks,
which appear naturally in GUT, extended Technicolor
or Compositeness models. 
The mixing of different multiplets by 
introducing a leptoquark--Higgs coupling would lead  to a contribution
to $0\nu\beta\beta$ decay
\cite{hir96a}. Combined with the half--life limit of the 
Heidelberg--Moscow experiment bounds on effective couplings can be derived
\cite{lepbb}. 
Assuming 
only one lepton number violating $\Delta L=2$ LQ--Higgs coupling unequal 
to zero and  the leptoquark masses 
not too different, one can derive from this limit either a bound on the 
LQ--Higgs coupling
\bea
Y_{LQ-Higgs}=(few)\cdot 10^{-6}
\eea
or a limit excluding leptoquarks with masses in the 
range of
${\cal O}(200 GeV)$. Assuming $Y_{LQ}\sim{\cal O}(1)$ leptoquark masses 
should be 
larger than (few) 10 TeV.

\section{Violations of the equivalence principle and Lorentz invariance}

Special relativity and the equivalence principle can be considered as the most 
basic foundations of the theory of gravity. However, string theories may
allow for or even predict the violation of these laws.
Such effects in the neutrino sector have been extensively studied in the 
framework of neutrino oscillations \cite{leung}.
A typical feature of the violation of Lorentz invariance (VLI) 
is that different
species of matter may have characteristic maximal attainable velocities.
The quantity $\delta v$ provides an observable for VLI. The corresponding
quantity describing violations of the equivalence principle (VEP) is the 
difference of characteristic couplings $\delta g$
to the gravitational potential $\phi$. While previous studies of neutrino 
oscillations are restricted to the region of large mixing of 
velocity/gravitational 
and flavor eigenstates, $0\nu\beta\beta$ decay provides a bound in the 
previously unconstrained region of zero mixing \cite{utp}: 
$\delta v<4 \cdot 10^{-16}$,
$\phi \delta g<4 \cdot 10^{-16}$.

\section{WIMP Dark Matter Search with Double Beta Experiments}

Weakly interacting masssive particles (WIMPs) such as the lightest 
supersymmetric
particle (LSP) are major candidates for the cold component of nonbaryonic
dark matter in the universe. 
Due to its low background properties double beta technology can also find 
applications in the search for direct detection of WIMPs. 
The Heidelberg--Moscow Experiment, without being specially designed
for this purpose,
gave the most stringent limits on WIMPs for several years \cite{beck}. 
New results with 0.69 kg y of measurement reached a background level of 0.042 
cts/(kg d keV) in the region between 15 keV and 40 keV. The derived limit
excludes WIMPS with masses greater than 13 GeV and cross sections as low as 
$1.12 \cdot 10 ^{-5}$ pb. These are the most stringent limits on 
spin-independent interactions using only raw data \cite{HM98}. 
The GENIUS experiment would allow to test almost 
the entire MSSM parameter space already in a first step using 
only 100 kg of enriched or even natural Ge \cite{Kla98,gen}. 

\section{Summary}
Neutrinoless double beta decay and dark matter search belong to the most 
sensitive approaches with great perspectives 
to test particle physics beyond the SM.
 
The possibilities to use $0\nu\beta\beta$ decay (and the most sensitive 
Heidelberg--Moscow experiment) for constraining 
neutrino masses, new interactions beyond the standard model, and violations of
Lorentz invariance and the equivalence principle
have been reviewed. Experimental 
limits on $0\nu\beta\beta$ decay are not only complementary to 
accelerator experiments, neutrino oscillations and cosmological precision 
measurements, but at least in some cases 
competitive or superior
to the best existing or planned approaches. 
Direct WIMP detection experiments 
can compete with recent and future accelerator 
experiments in the search for SUSY and experiments using double beta 
technology belong to the most promising approaches in this field of research.
A further large breakthrough, 
both for double beta decay and dark matter search, 
will be possible realizing the GENIUS proposal, which would improve the obtained limits by up to 1-2 orders of magnitude.

\end{document}